\documentclass[%
 reprint,
 amsmath,amssymb,
 aps,
]{revtex4-1}

\usepackage{graphicx}
\usepackage{dcolumn}
\usepackage{bm}

\begin{document}

\preprint{APS/123-QED}

\title{Static wetting on deformable substrates, from liquids to soft solids}

\author{Robert W. Style}
\affiliation{%
Department of Geology \& Geophysics, Yale University, New Haven, CT 06520, USA. 
}%
 \altaffiliation[Also at ]{Department of Mechanical Engineering and Material Science, Yale University, New Haven, CT 06520, USA} 
\author{Eric R. Dufresne}%
\affiliation{%
Department of Mechanical Engineering and Material Science, Yale University, New Haven, CT 06520, USA. 
}%

%

\date{\today}

\begin{abstract}
Young's law fails on soft solid and liquid substrates where there are substantial deformations near the contact line.
On liquid substrates, this is captured by Neumann's classic analysis, which provides a  geometrical construction for minimising the interfacial free energy.
On soft  solids, the total free energy includes an additional contribution from elasticity.
A linear-elastic model incorporating an out-of-plane restoring force due to solid surface tension was recently shown to accurately predict the equilibrium shape of a thin elastic film due to a large sessile droplet.
Here, we extend this model to find substrate deformations due to droplets of arbitrary size. 
While the macroscopic contact angle matches Young's law for large droplets, it matches Neumann's prediction for small droplets.
The cross-over droplet size is roughly given by the ratio of the solid's surface tension and elastic modulus. 
At this cross-over, the macroscopic contact angle increases, indicating that the substrate is effectively less wetting. For droplets of all sizes, the microscopic behaviour near the contact line follows the Neumann construction giving local force balance.
\end{abstract}

\pacs{Valid PACS appear here}
\maketitle


\section{Introduction}

\begin{figure}[h]
\centering
  \includegraphics[width=9cm]{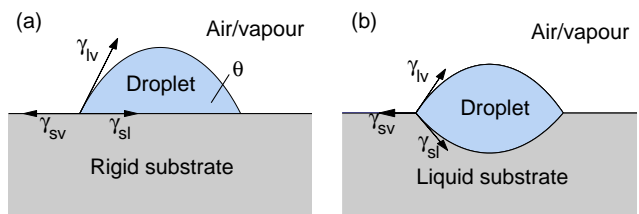}
  \caption{Wetting on (a) rigid and (b) liquid substrates as described by Young and Neumann, respectively.}
  \label{fig:schem}
\end{figure}

Wetting is a fundamental physical process with far-ranging applications \cite{dege10,butt03,quer08,kuma07}. 
In the absence of long-range interfacial forces, our understanding of wetting typically centres upon two key results: Young's law for wetting on rigid substrates and Neumann's triangle for wetting on liquid substrates \cite{youn05,neum94,dege10,mich62}. 
These are shown schematically in Figure \ref{fig:schem}. 
On a rigid substrate, the total free energy is minimized when  $\gamma_{sv}=\gamma_{sl}+\gamma_{lv}\cos\theta$.  
Here,  $\gamma_{sl}$, $\gamma_{sv}$ and $\gamma_{lv}$, are the intensive free energies for the solid-liquid, solid-vapour and liquid-vapour interfaces, respectively \cite{dege10}.
This result is often visualized in terms of an in-plane balance of surface tensions, whose magnitudes are given by the interfacial energies, acting at the contact line.
This interpretation of Young's law is not rigorously correct since solid surfaces can have contibutions to the surface tension  beyond the interfacial energy  \cite{shut50}.
On a liquid substrate, the three interfaces arrange so that surface tensions balance in- and out-of-plane at the contact line.
This configuration, called Neumann's triangle, uniquely determines the angles between the phases at the contact line\cite{neum94,lest61}. 
For liquids, the equivalence of surface tensions and interfacial energies is not problematic.
For small drops, where the effect of gravity is unimportant, droplets on a liquid substrate then assume lenticular forms, such as that shown in Figure \ref{fig:schem}(b), where the interfaces between phases take the form of spherical caps\cite{dege10}. 

Young's law and Neumann's triangle are useful in explaining many wetting phenomena. 
However, in a sense they are two extremes in a continuum of wetting problems: Young's law gives the behaviour in the limit of an infinitely hard substrate, while Neumann's triangle gives the result for an infinitely soft (\emph{i.e.} fluid) substrate. 
Wetting on soft solid substrates should then exhibit a spectrum of intermediate behaviour. 
This liquid-solid crossover is suggested  by several recent results. 
Experiment, analytical theory and molecular-dynamics simulations, suggest that the shape of a soft substrate near a contact line does not depend on the substrate elastic modulus \cite{jeri11,leon11}. 
Eslami and Elliott\cite{esla11} showed that observations of condensation on soft substrates can be explained by treating soft surfaces as fluids. 
Mora \emph{et al.} \cite{mora10} showed that the Rayleigh-Plateau instability can occur in thin elastic strands. 

In this paper, we generalize Jerison \emph{et al.}'s solution for a large droplet on a thin solid substrate  to droplets of arbitrary radius. 
For large droplets, we recover Young's law for the macroscopic contact angle.
For small droplets, the macroscopic behaviour is given by a lenticular shape, such as that seen in Figure \ref{fig:schem}(b), as predicted by Neumann's analysis for wetting on liquid substrates.  
The cross-over droplet size is approximately the ratio of the solid's surface tension and elastic modulus.
For droplets of all sizes, the microscopic behaviour near the contact line follows a Neumann construction so as to give local force balance.

\section{Deformation of an Elastic Slab with Surface Tension \label{sec:theory}}

Recently, Jerison \emph{et al.} \cite{jeri11} measured the  deformation of a thin film of soft silicone gel under a large water droplet. 
Their results showed that the surface profile was well-matched by a linear-elastic theory that included the out-of-plane restoring force due to the substrate's  surface tension. 
The theory they presented was two-dimensional, and thus applicable to very large droplets. 
However in order to apply the theory to droplets of a finite radius, we need to extend their results to three dimensions. To this end, we solve the elastic equations in cylindrical coordinates for a uniform substrate of thickness $h$ and of infinite horizontal extent.

\subsection{Purely elastic deformations}

We take a substrate which is pinned to a rigid surface at $z=0$ with displacements there being zero. 
In cylindrical-polar coordinates, the governing equations are the steady-state Navier equations
\begin{equation}
(1-2\nu)\left(\nabla^2u_r-\frac{u_r}{r^2}\right)+\frac{\partial}{\partial r}(\nabla.\mathbf{u})=0
\nonumber
\end{equation}
and
\begin{equation}
(1-2\nu)\nabla^2u_z+\frac{\partial}{\partial z}(\nabla.\mathbf{u})=0
\end{equation}
where we have assumed cylindrical symmetry, $u_r$ and $u_z$ are the $r$ and $z$ components of the displacement $\mathbf{u}$, and $\nu$ is the Poisson's ratio\cite{irge08}.
The stress is related to the displacements by
\begin{equation}
\label{eqn:stress_disp}
\sigma_{ij}=\frac{E}{1+\nu}\left[\frac{1}{2} \frac{\partial u_i}{\partial x_j}+\frac{1}{2}\frac{\partial u_j}{\partial x_i}+\frac{\nu}{1-2\nu}\frac{\partial u_k}{x_k}\delta_{ij} \right],
\end{equation}
where $E$ is Young's modulus.
Because of the radial symmetry, we take Hankel transforms of the stress and displacement fields, giving $\hat{u}_r(s,z)=\mathbf{H}_1[u_r(r,z)]$, $\hat{u}_z(s,z)=\mathbf{H}_0[u_z(r,z)]$, $\hat{\sigma}_{rz}(s,z)=\mathbf{H}_1[\sigma_{rz}(r,z)]$ and $\hat{\sigma}_{zz}(s,z)=\mathbf{H}_0[\sigma_{zz}(r,z)]$. Here $\mathbf{H}_0$ and $\mathbf{H}_1$ are Hankel transforms of order 0 and 1 respectively\cite{sned51}. By following the same solution method used Xu \emph{et al.} \cite{xu10}, we then obtain
\begin{equation}
\hat{\sigma}_i(s,h)=Q_{ij}(s,h,z) \hat{u}_{j}(s,z),
\label{eqn:sigmaQu}
\end{equation} 
where $\hat{\sigma}_i(s,z)=(\sigma_{rz}(s,z),\sigma_{zz}(s,z))$ and $\hat{u}_i(s,z)=(\hat{u}_r(s,z),\hat{u}_z(s,z))$.
The spring constant $Q_{ij}(s,h,z)$ that relates surface tractions to substrate displacement is given by
\begin{equation}
Q_{ij}(s,h,z)=\frac{E}{1+\nu}P_{ij}(s,h)M^{-1}_{ij}(s,z)
\end{equation}
where 
\begin{flalign}
P_{ij}(s,h)=\left(\begin{array}{cc}
0 & -\frac{s}{2} \\
\frac{s\nu}{1-2\nu} & 0 \end{array} \right)&M_{ij}(s,h)
+  \nonumber \\
&\left( \begin{array}{cc}
\frac{1}{2} & 0 \\
0 & \frac{1-\nu}{1-2\nu} \end{array}\right) \left.\frac{\partial M_{ij}(s,z)}{\partial z}\right|_{z=h}
\end{flalign}

and
\begin{flalign}
&M_{ij}(s,z)=  \nonumber \\
&\left(
\begin{array}{cc}
 \frac{(3-4\nu) \sinh (zs)+zs \cosh (zs)}{4 s(1- \nu)} & \frac{z \sinh
   (zs)}{2(1-2 \nu) } \\
 \frac{-z \sinh (zs)}{4 (1-\nu)} & \frac{(3-4\nu) \sinh (zs)-zs \cosh (zs)}{2s(1-2 \nu)} \\
\end{array}
\right).
\end{flalign}

Knowing $Q_{ij}$ enables the calculation of displacements throughout a substrate given forces applied to its surface, as it is directly related to the Green's function for the problem.
For example, in the case of purely vertical forces acting on the surface, the surface displacements are given by $u_z(r,h)=\mathbf{H}^{-1}_0[Q_{zz}^{-1}(s,h,h)\hat{\sigma}_{zz}(s,h)]$, where
\begin{flalign}
&Q^{-1}_{zz}(s,h,h)= \nonumber \\
&\frac{2(1-\nu^2)}{sE}\frac{(3-4\nu)\sinh(2sh)-2sh}{5-12\nu+8\nu^2+2s^2h^2+(3-4\nu)\cosh(2sh)}.
\label{eqn:Qinv}
\end{flalign}
As a useful check, $Q^{-1}_{zz}(s,\infty,\infty)=2(1-\nu^2)/(sE)$ which gives Terezawa's displacement solution for axisymmetric forces acting on a semi-infinite elastic substrate \cite{sned51}.

\subsection{Elasticity with surface stresses}

The elastic response due to stresses applied at the free surface is given by Equation (\ref{eqn:sigmaQu}). 
However, to describe the deformation due to the wetting of a liquid droplet, we also need to include surface stresses -- a  generalization of surface tension for solids. 
The surface stress, $\Upsilon$, is related to the surface energy $\gamma$ by the Shuttleworth equation
\begin{equation}
\label{eqn:shut}
\Upsilon_{ij}=\gamma \delta_{ij}+\frac{\partial \gamma}{\partial \epsilon_{ij}},
\end{equation}
where $\delta_{ij}$ is the Kronecker delta, and $\epsilon_{ij}$ is the surface strain \cite{shut50,butt03}. $\Upsilon$ is generally anisotropic, however it does simplify under certain situations.
For instance at a fluid-fluid interface $\frac{\partial \gamma}{\partial \epsilon_{ij}}=0$, so the surface stress is isotropic and equal to the surface energy. 
Both $\Upsilon$ and $\gamma$ can then be called the `surface tension' without ambiguity.
For many isotropic solid materials the surface stresses are also approximately isotropic \cite{peth57,camm94,spae00}, and $\Upsilon$ and $\gamma$ are typically of similar magnitudes \cite{camm94}.
However they are not necessarily equal \cite{spae00,marc12} and we have to distinguish carefully between them.
Here we follow Shuttleworth \cite{shut50} in referring to $\Upsilon$ as the `surface tension' with the understanding that it represents the actual tension force at the surface of the solid.

For tractability, we assume isotropic surface tensions, and that the solid-vapour surface tension $\Upsilon_{sv}$ and the solid-liquid surface tension $\Upsilon_{sl}$ are the same and given by $\Upsilon_s$.
Later we discuss expected changes in more general cases.

The linearised surface tension force is $\mathbf{\sigma}_\Upsilon=\Upsilon_s\frac{1}{r}\frac{\partial}{\partial r} \left(r\frac{\partial u_z(r,h)}{\partial r}\right)\mathbf{\hat{z}}$, where $\mathbf{\hat{z}}$ is the unit vector in the vertical direction \cite{camm94}. We include this in the force balance at the surface \cite{jeri11} to find that 
\begin{equation}
\label{eqn:spring_const_with_surf_tension}
\hat{\sigma}_i(s,h)=QS_{ij}(s,h,z) \hat{u}_{j}(s,z),
\end{equation} 
where
\begin{equation}
QS_{ij}(s,h,z)=Q_{ij}(s,h,z)+ \left( \begin{array}{cc}
0 & 0 \\
0 & \Upsilon_s s^2 \end{array}\right).
\end{equation}
In particular, we have
\begin{flalign}
&QS^{-1}_{zz}(s,h,h)= \nonumber \\
&\frac{2(1-\nu^2)}{sE}\frac{1}{\frac{5-12\nu+8\nu^2+2s^2h^2+(3-4\nu)\cosh(2sh)}{(3-4\nu)\sinh(2sh)-2sh}+\frac{2(1-\nu^2)s\Upsilon_s}{E}}.
\label{eqn:QSinv}
\end{flalign}
%

$QS^{-1}_{zz}(s,h,h)$ tells us how the surface of the substrate responds to imposed forcings of wavelength $O(1/s)$, and we can use this to gain some interesting insights into the physics of the problem. For small wavenumber$s$, $QS^{-1}(s,h,h)\rightarrow Q^{-1}(s,h,h)$.  That is, for long-wavelength surface perturbations, the response of the substrate is purely elastic. 
On the other hand, for large $s$, $QS^{-1}(s,h,h)\rightarrow 1/(\Upsilon_s s^2)$.   
That is, for short wavelength perturbations to the surface, the force due to surface tension  dominates the  substrate response. 
Analysis of Equation (\ref{eqn:QSinv}) shows us that the cross-over length scale is $\Upsilon_s/E$, provided the substrate thickness $h\gtrsim \Upsilon_s/E$.
Perturbations of lengthscale $\lambda\gg \Upsilon_s/E$ are damped elastically, while perturbations of length scale $\lambda \ll \Upsilon_s/E$ are damped by surface tension\cite{long96}. For thin substrates $h<\Upsilon_s/E$, the substrate is less compliant due to the presence of the rigid bottom boundary, so the elastic response to perturbations is stronger, and the crossover lengthscale is reduced. As we shall see, the length scale $\Upsilon_s/E$ repeatedly emerges in substrate micro-deformation problems as a controlling influence on the behavioural response of the substrate. Because it represents the balance between elasticity and capillarity we refer to it here as the elasto-capillary length, though it should be noted that there are several other length scales that have also been given this title\cite{roma10}.

At this point it is worth correcting a small error in the analysis of Jerison \emph{et al.} \cite{jeri11}. 
When including the surface tension of the substrate in their model, they used $\mathbf{\sigma}_\Upsilon=\frac{\Upsilon_s}{2}\frac{d^2u_z(x,h)}{dx^2}$. 
The correct linearised surface tension force does not have the factor of $1/2$. This can be derived from the fact that for a curved surface $\mathbf{\sigma}_\Upsilon=\Upsilon_s {\cal K} \mathbf{n}$ where ${\cal K}$ is the surface curvature and $\mathbf{n}$ is the normal vector to the surface\cite{shut50}. 
The surface-tension force is the linearised version of this expression. 
Importantly this correction does not affect  Jerison \emph{et al.}'s claim that that a linearised elastic model matches the observed surface profile.  
Rather,  it changes the fitted ratio of the liquid and solid surface tensions.

\section{Solution for a hemispherical droplet}

Using the results above, we can calculate the exact deformation of the surface caused by a hemispherical droplet of radius $R$. Assuming that all length scales in the problem are much bigger than the typical range of intermolecular forces, then the surface tension of the droplet appears as a line force, and the traction imposed by the droplet on the surface of the substrate is given by
\begin{equation}
\mathbf{T}(r,h)=\gamma_l \delta(r-R)\mathbf{\hat{z}} - P_lH(R-r)\mathbf{\hat{z}}
\end{equation}
where $\gamma_l$ is the liquid/vapour surface tension, $P_l$ is the Laplace pressure in the droplet, $\delta(x)$ is the Dirac delta function, $H(x)$ is the Heaviside step function, and $ \mathbf{\hat{z}}$ is the unit vector in the $z$ direction \cite{shan86}.
The first term corresponds to the out-of-plane surface tension force of the droplet at the contact line.
Here we have assumed that the liquid-vapor interface is oriented normal to the substrate in accordance with Young's law.
We will relax this constraint in Section IV.
The second term corresponds to the Laplace pressure force. 
To ensure mechanical equilibrium, the total force exerted by the droplet on the substrate is zero, thus  the distributed Laplace pressure inside the droplet must balance the localised force of the droplet surface tension at the contact line\cite{dege10}. For our hemispherical droplet, this means $P_l=2\gamma_l/R$. 
Taking the Hankel transform $\mathbf{H}_0[\mathbf{T}(r,h)]$, we find that $\hat{\sigma}_{zz}(s,h)=\gamma_l R J_0(sR)-2\gamma_l J_1(sR)/s$, where $J_i(z)$ is the $i$th order Bessel function of the first kind. Therefore, using Equation (\ref{eqn:spring_const_with_surf_tension}) and taking the inverse Hankel transform we find the vertical surface deformation  of the substrate that results from the presence of the droplet:
\begin{flalign}
&u_z(r,z)= \nonumber \\
&\gamma_l\int_0^\infty s\left[RJ_0(sR)-2\frac{J_1(sR)}{s}\right]QS^{-1}_{zz}(s,h,z) J_0(sr)ds.
\label{eqn:uz}
\end{flalign}

The horizontal displacements $u_r(r,z)$ can be similarly calculated.

\subsection{Large droplets on a thin substrate: $R\gg h$ \label{sec:thin_substrate}}

There are three asymptotic limits that are of particular use. Firstly, when the droplet is very large compared to the substrate thickness, we expect that the peak profile should approach the two-dimensional solution given by Jerison \emph{et al.} \cite{jeri11}. 
If we define $\epsilon=h/R$, nondimensionalise $s$ by setting $\bar{s}=sh$, and move into the frame of reference of the peak by setting $\bar{x}=(r-R)/h$. 
Then we can expand the results for $u_z$ and $u_r$ from above in powers of $\epsilon$, to find that at leading order the surface displacements are
\begin{widetext}
\begin{eqnarray}
u_z(\bar{x})= 
\frac{2\gamma_l(1-\nu^2)}{\pi E}\int_0^\infty \frac{\cos(\bar{s}\bar{x})d\bar{s}}{\frac{5-12\nu+8\nu^2+2\bar{s}^2+(3-4\nu)\cosh(2\bar{s})}{(3-4\nu)\sinh(2\bar{s})- 2\bar{s} }\bar{s}+\frac{2(1-\nu^2)\Upsilon_s}{Eh}\bar{s}^2},\label{eqn:R_gg_h}\\
u_r(\bar{x})=\frac{\gamma_l(1+\nu)}{\pi E}\int_0^\infty \frac{\left(\frac{-3-2\bar{s}^2+10\nu-8\nu^2+(3-10\nu+8\nu^2)\cosh(2\bar{s})}{(3-4\nu)\sinh(2\bar{s})-2\bar{s}}\right)\sin(\bar{s}\bar{x})d\bar{s}}{\frac{5-12\nu+8\nu^2+2\bar{s}^2+(3-4\nu)\cosh(2\bar{s})}{(3-4\nu)\sinh(2\bar{s})- 2\bar{s} }\bar{s}+\frac{2(1-\nu^2)\Upsilon_s}{Eh}\bar{s}^2}\label{eqn:R_gg_h2}
\end{eqnarray}
\end{widetext}
These are the same as the results given by Jerison \emph{et al.} \cite{jeri11}, after the correction of a small error in their analysis noted above. 
Thus we recover the two-dimensional solution. 

\begin{figure}[h]
\centering
  \includegraphics[width=9cm]{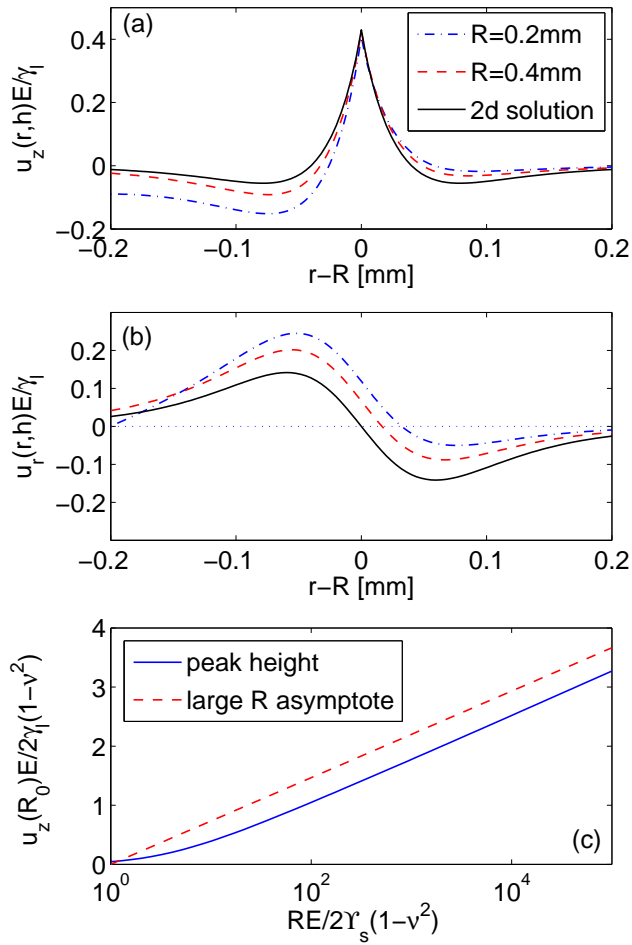}
  \caption{Substrate deformation with droplet size  (a) out-of-plane displacements,  and (b) in-plane displacements. The drop is to the left of the wetting ridge. $h=50\mu$m, $E=3$kPa, $\Upsilon_s=0.06$N/m and $\nu=1/2$. For $R=0.2$ mm (blue dashed-dotted curve) and $0.4$ mm (red dashed curve), the profile is computed using Equation (\ref{eqn:uz}) and its $u_r$ analogue.  The large droplet limit is computed using Equations (\ref{eqn:R_gg_h}) and (\ref{eqn:R_gg_h2}) (black curve).   (c) The height of the wetting ridge on a semi-infinite substrate. The dashed line shows the asymptote for $RE/\Upsilon_s\gg1$.}
  \label{fig:asymptotics}
\end{figure}

Figures \ref{fig:asymptotics}(a) and (b) demonstrate how the vertical and horizontal surface displacements tend to the two-dimensional solution as the droplet diameter increases. 
For smaller droplets, the shape of the wetting ridge is asymmetric, as has been observed experimentally\cite{peri08}. As the droplet radius increases the shape of the wetting ridge becomes more symmetric.
For $R/h\gtrsim {\cal O}(10)$, the two-dimensional solution provides a good estimate to $u_z(r,h)$ near the contact line. However a significant difference persists between $u_r(r,h)$ and the two-dimensional theory up to relatively large droplet radii. This can explain the deviation between experiments and theory found by Jerison \emph{et al.} \cite{jeri11}. They measured droplets with $R/h\sim {\cal O}(10)$ and found excellent agreement between the two-dimensional theory and experiments for vertical displacements. However experimentally-observed horizontal displacements were asymmetric and shifted towards positive displacements relative to theoretical predictions. Our results suggest that this is simply due to the finite droplet size.


\subsection{Small droplets: $R\ll h$, $R \ll \Upsilon_s/E$ \label{sec:small_droplets}}
Consider a sessile droplet of radius $R\ll \Upsilon_s/E$ on a substrate of thickness $h\gg R$. 
Since the  wavelength of applied surface stresses can be no bigger than than the droplet radius, then the arguments from the end of Section \ref{sec:theory} imply that  surface tension dominates the response of the substrate.  
As elasticity effectively drops out of the problem at these small lengthscales, the system is the same as that of a droplet of liquid sitting at the interface between two other fluids. 
Thus, the solution must simply be that of Neumann's classic three-fluid contact problem shown in Figure \ref{fig:schem}(b), with $\gamma_{sl}$ and $\gamma_{sv}$ replaced by $\Upsilon_s$.
Here, the droplet takes a lenticular shape formed by the union of  two  spherical caps, with  contact angles  given by Neumann's triangle \cite{lest61}.

\subsection{Droplets on a semi-infinite substrate: $R\ll h$, $\Upsilon_s/E\ll h$}
In the case of an infinitely-thick substrate, as $sh\rightarrow\infty$, we find that Equation (\ref{eqn:QSinv}) simplifies considerably. We nondimensionalise using the droplet radius as a length scale, so that $\tilde{r}=r/R$ and $\tilde{s}=sR$ to find that the surface displacement is
\begin{equation}
\label{eqn:inf_thick}
u_z(\tilde{r}R)=\frac{2\gamma_l(1-\nu^2)}{ E}\int_0^\infty \frac{\left[J_0(\tilde{s})-2\frac{J_1(\tilde{s})}{\tilde{s}}\right]J_0(\tilde{s}\tilde{r})d\tilde{s}}{1+\frac{2(1-\nu^2)\Upsilon_s}{RE}\tilde{s}}.
\end{equation}

	This equation shows two interesting details. 
Firstly, the height of the peak has a linear dependence on $\gamma_l/E$, as predicted by many theoretical works \cite{carr96}, but is also a weak function of $RE/\Upsilon_s$. 
Asymptotically we find that for $R\gg \Upsilon_s/E$,
\begin{equation}
 \label{eqn:asy_thick_substr}
 u_z(R)\sim \frac{2\gamma_l(1-\nu^2)}{E\pi}\log\left(\frac{RE}{2(1-\nu^2)\Upsilon_s}\right) + \mathrm{const.}
\end{equation}
Figure \ref{fig:asymptotics}(b) shows the peak height as a function of droplet radius, along with this leading order asymptotic result without the constant term. 
Evidently, the asympotic expression is a convenient upper bound on the surface displacement of a substrate caused by a drop of radius $R$.
This is in contrast to Terezawa's solution for the case of a circular line force on a semi-infinite substrate, where the substrate strain diverges at the contact line\cite{sned51}, and the two-dimensional models of Jerison \emph{et al.}\cite{jeri11} and Long \emph{et al.}\cite{long96,peri09} that predict a divergent peak height for semi-infinite substrates.
Our asymptotic expression resembles the logarithmic divergence of peak height with $R$ found by White \cite{whit03}.
 However, while White's expression depends on the range of intermolecular forces,  our result only depends on drop radius and the elastocapillary lengths $\Upsilon_s/E$ and $\gamma_l/E$. 

\section{Discussion}

\subsection{Does Young's law hold on soft substrates? \label{sec:young}}

Above, we have imposed a particular macroscopic contact angle $\theta$ for the droplet, namely the value given by Young's law. 
However, some experiments have suggested, this may not be the equilibrium contact angle on a soft substrate \cite{soon86}. 
Therefore, we seek to determine the range of validity of  Young's law using the model derived above. 

\begin{figure*}
\centering
  \includegraphics[height=6cm]{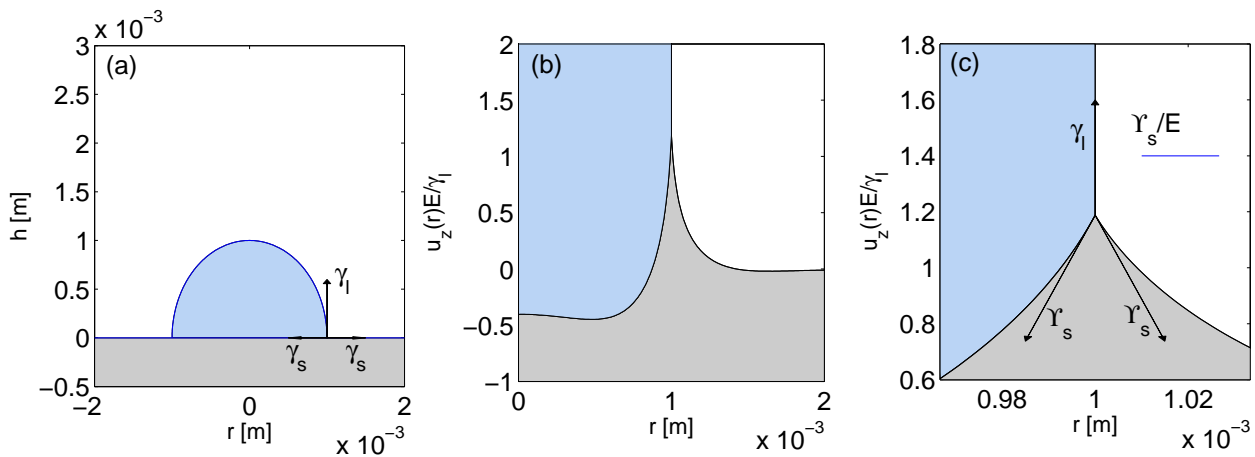}
  \caption{Substrate deformation for a hemispherical 1 mm radius droplet of water on an incompressible elastic substrate of thickness $h=0.5$mm with $\Upsilon_s=0.05$N$/$m and $E=3$kPa. (a) Macroscopic view. (b) Substrate displacement underneath the droplet. Here,  the vertical axis is scaled by $\gamma_l/E$. (c) Close up of the tip of the wetting ridge, with Neumann's triangle superimposed.}
  \label{fig:drop_diff_scales}
\end{figure*}


We calculate the equilibrium shape of a sessile droplet on a soft substrate by minimizing its free energy, written as $
F=F_{surf}+F_{el}$, where the contribution from surface energies,
\begin{equation}
F_{surf}=\gamma_l A_{lv}+(\gamma_{sl}-\gamma_{sv})A_{sl},
\end{equation} with $A_{lv}$ being the area of the liquid-vapour interface, and $A_{sl}$ being the area of the substrate-liquid interface.  Note that we use the surface energies $\gamma_{sl}$ and $\gamma_{sv}$ here rather than the surface stresses.
The elastic contribution is
\begin{equation}
F_{el}=\frac{1}{2}\int_{A_{sl}} \mathbf{T}.\mathbf{u} dA.
\end{equation}
Note that we ignore the contribution to $F$ of the line tension $\tau$. This is only expected to be significant when the size of a droplet $\sim \tau/\gamma_l$ \cite{dege10}. Typically $\tau\sim 10^{-11}$J/m \cite{dege10}, so line tension can generally be ignored for droplets larger than a nanometre.

The contact angle for a droplet is then found by minimising this energy subject to the constraint that the total droplet volume is constant. The elastic energy starts to impact the equilibrium contact angle when $F_{el}\sim F_{surf}$ and we can predict when this will occur by noting that for our hemispherical droplet,
\begin{equation}
\label{eqn:Fel}
F_{el}=\pi R\gamma_{l}u_z(R,h)-\frac{2\pi\gamma_l}{R}\int_0^{R}ru_z(r,h)dr.
\end{equation}
The first term corresponds to the work done by the surface tension of the droplet in pulling up the wetting ridge, while the second term corresponds to the work done by the Laplace pressure of the droplet. Each of these two expressions scale like $\gamma_l R u_z(R,h)$, and so using Equation (\ref{eqn:asy_thick_substr}) gives an upper bound on the magnitude of the elastic energy: $F_{el}\lesssim \frac{R\gamma_l^2}{E}\log(RE/\Upsilon_s)$. $F_{surf}\sim\gamma_l R^2$, so we find that the ratio $F_{el}/F_{surf}\sim (\gamma_l/RE) \log(\Upsilon_s/RE)$. This is small provided that $R\gg \Upsilon_s/E,\gamma_l/E$, in which case the deformation of the substrate will have little influence on the energetics of the droplet. Then the contact angle will be the same as it is for a rigid substrate. On the other hand, as the droplet radius reduces towards $\mathrm{max}[\Upsilon_s/E,\gamma_l/E]$, the growing influence of the substrate deformation will start to manifest itself as a change in the macroscopic contact angle.

The key result here is that Young's law is recovered for droplets that are much larger than the two elasto-capillary lengths $\Upsilon_s/E$ and $\gamma_l/E$,  \emph{i.e.} $\gamma_{sv}=\gamma_{sl}+\gamma_{lv}\cos\theta$. An instructive example of this is given in Figure \ref{fig:drop_diff_scales}(a). In this case, $R \gg \Upsilon_s/E, \gamma_l/E$. Therefore, macroscopically the droplet conforms to Young's law, with $\theta=90^\circ$ as we assume $\gamma_{sl}=\gamma_{sv}$. However if we zoom in on the contact line, we see that the local angle between the liquid-air and substrate liquid interfaces deviates substantially from $\theta$. This is clearly demonstrated in Figure \ref{fig:drop_diff_scales}(b). Note that this second figure is calculated using Equation (\ref{eqn:uz}) and the vertical length scale is non-dimensionalised with $\gamma_l/E$.

\begin{figure}[h]
\centering
  \includegraphics[height=10cm]{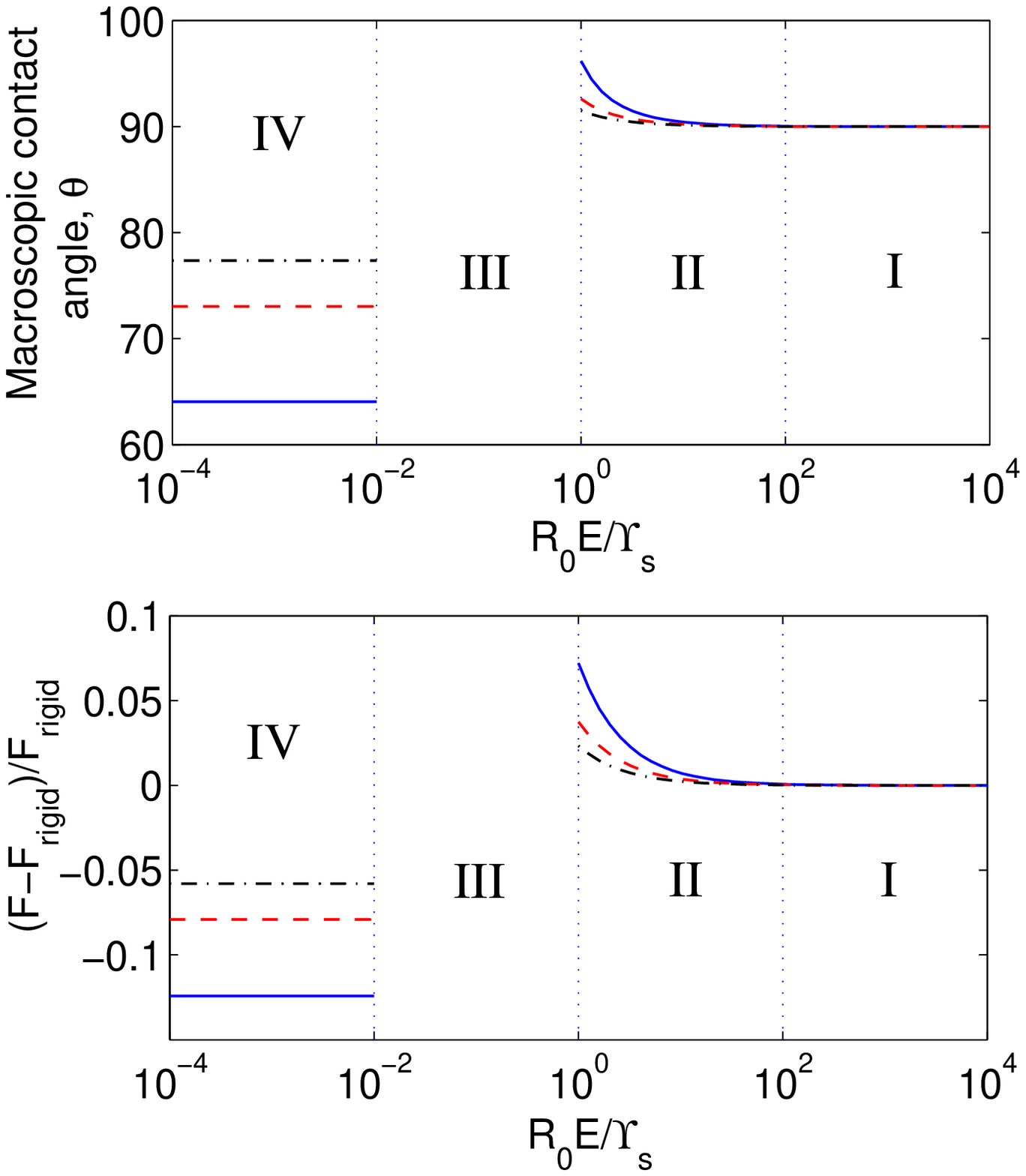}
  \caption{Effect of substrate deformability on (a) contact angle and (b) adsorption free energy. Change in free energy relative to the free energy of a droplet on a rigid substrate $(F-F_{rig})/F_{rig}$. For both plots, $\gamma_l=0.07$N/m, $E=3000$Pa, $h=20\mu$m. The different lines correspond to different values of $\Upsilon_s$; continuous line: $\Upsilon_s=0.08$N/m, dashed line: $\Upsilon_s=1.2$N/m, dash-dotted line: $\Upsilon_s=1.6$N/m. Roman numerials indicate  different regimes of behaviour as described in the text.}
  \label{fig:contact_angle_free_energy}
\end{figure}

In order to demonstrate how Young's law holds for large droplets, and also how the contact angle changes as $R$ approaches $\Upsilon_s/E$ and $\gamma_l/E$, we calculate the change in the macroscopic contact angle, $\theta$, for an incompressible, thin substrate. 
The details of this calculation are given in Appendix A. Figure \ref{fig:contact_angle_free_energy}(a) shows the macroscopic contact angle for a droplet of volume $V=4/3\pi R_0^3$, with $E=3$kPa, $h=20\mu$m and $\gamma_l=0.07$N/m. $\gamma_{sl}=\gamma_{sv}$ so the macroscopic contact angle given by Young's law is $90^\circ$. We plot the curve for $\theta$ for $\Upsilon_s=0.08,$ 0.12 and 0.16N/m. There are four different regimes of behaviour. In regime I, where $R\gg \Upsilon_s/E,h$, Young's law holds in agreement with the analysis above. In regime II, as the size of the droplet shrinks towards $\Upsilon_s/E$ (though $R\gg h$), $\theta$ starts to increase. This occurs because the elastic energy required to deform the substrate into a wetting ridge acts like a line tension, causing the droplet to attempt to reduce its wetted surface area. This makes the substrate appear less wetting. In regime IV, when the droplet becomes much smaller and $R_0\ll \Upsilon_s/E,h$ then, as we showed in Section \ref{sec:small_droplets}, the substrate responds like a fluid and the system behaves like the classical three-fluid problem shown in Figure \ref{fig:schem}(b). Thus $\theta$ reduces to the value given by Neumann's triangle, with substrate surface tensions replaced by surface stresses. Unfortunately, the linear constraints of the model, detailed further in Appendix A, mean that it is difficult to estimate the contact angle behaviour in regime III. However it is interesting to note that there must be a pronounced maximum in $\theta$ as $R\rightarrow 0$. Note that we do not assume that $\Upsilon_s=\gamma_s$ anywhere in this calculation.

As well as showing the change in contact angle with droplet size in Figure \ref{fig:contact_angle_free_energy}, we also show the free energy of droplet adhesion, $F$, changes with $R_0$. This is normalised by the free energy of adhesion for a rigid substrate $F_{rigid}$. In regime II, there is an increase in the free energy above that of a rigid substrate, while in regime IV the free energy drops to a lower energy state than that in the rigid-substrate case.

Each of the regimes above should be  observable experimentally. 
For typical liquid droplets on gels (with kPa-scale moduli), the elastocapillary length is in the micron scale.
Thus, droplets in regime II should be readily observable with light microscopy.
Droplets in regime IV could be readily generated by condensation and should be observable with atomic force microscopy. 
In that case, our results suggest that condensation will occur more rapidly onto a soft substrate than onto a rigid substrate due to the lower energy state of droplets on the softer substrate, as shown in Fig. \ref{fig:contact_angle_free_energy}(b). 
This is in agreement with the experiments of Sokuler \emph{et al.}\cite{soku10} and the theory of Eslami and Elliott\cite{esla11}.
By contrast, the elastocapillary length for soft elastomers and rubbers ($E\sim1$MPa) is $\sim50$nm, so regime I and II drops will be achievable, but interfacial forces will likely be important in regimes III and IV, which will lead to other physics being important in the problem.
Finally the elastocapillary length for hard materials such as glass (GPa and above) is at, or below, molecular dimensions, so any droplet will be in regime I. 

It is worth briefly noting two points. Firstly, our results also indicate that the predicted deviations in contact angle from Young's law also depend upon the size of the solid surface tension $\Upsilon_s$, as shown in Figure \ref{fig:contact_angle_free_energy}(a). For larger values of $\Upsilon_s/\gamma_l$, the changes in contact angle become small, while when $\Upsilon_s/\gamma_l$ is smaller, there can be quite significant changes in contact angle, especially when $R_0E/\Upsilon_s\ll 1$. Physically this occurs because when $\Upsilon_s/\gamma_l$ is large, the liquid surface tension cannot overcome the solid surface tension. This means that the system approximates that of a droplet on a rigid interface, and so the contact angle approaches that given by Young's law. On the other hand, as $\Upsilon_s/\gamma_l$ reduces, the stronger liquid surface tension causes surface deformations to increase, resulting in larger changes in $\theta$. Secondly, we note that our predictions differ from those derived theoretically by Shanahan \cite{shan87} and Leonforte and Muller \cite{leon11}. Shanahan \cite{shan87} considered a local energy minimisation at the contact line in his derivation of the contact angle. This is in contrast to the global energy minimisation we perform to calculate $\theta$.
Leonforte and Muller primarily considered nano droplets including intermolecular forces and line tension.
They also performed a scaling analysis to approximate the competition of surface tension and elasticity and concluded that for very small droplets Young's law would hold, while for larger droplets Young's law would break down.  


\subsection{Force balance at the contact line \label{sec:neumann}}

While Young's law holds for the macroscopic contact angle for droplets with $R \gg \gamma_l/E,\Upsilon_s/E$, the microscopic behavior at the contact line is quite different.
Figure \ref{fig:drop_diff_scales} shows the equilbrium of the three-phase system across length scales.
Near the three-phase contact line ($|r-R|\ll \Upsilon_s/E,R$),  the wetting ridge forms a cusp with a well defined angle, as shown in Figure \ref{fig:drop_diff_scales}(c).   
Interestingly, this angle is identical to the one predicted by Neumann's triangle with solid surface tensions $\Upsilon_s$, so local force balance between surface tensions is enforced at the tip of the ridge.
An analytical proof of this result is given in Appendix B.
This also means that the contact between the three phases behaves much like the contact between three fluids. 
As $E\rightarrow 0$, the size of this fluid-like region grows, and the substrate behaviour approaches a completely fluid-like response. 
On the other hand, as $E$ becomes large, the ridge height becomes small, as can be seen from Equation (\ref{eqn:asy_thick_substr}), and the response of the substrate returns to the flat, rigid surface of Young's law.
Note that this picture starts to break down when the height of the peak is comparable  to the range of interfacial forces ($\sim$2-10\AA), which will effect the local behaviour at the contact line\cite{whit03,das11}.

Importantly, the presence of a small region where interfaces obey Neumann's triangle at the contact line ensures balance in the out-of-plane component ignored by Young's construction. 
This resolves the apparent lack of force balance in Figure \ref{fig:schem}(a) \cite{maxw78,oliv10}.

\section{Conclusions}
We consider a sessile droplet placed on a soft substrate.
We find that large droplets satisfy Young's law for wetting on solid substrates, while small droplets satisfy Neumann's construction for wetting on liquid substrates.
The cross-over size is given by the elasto-capillary lengths $\Upsilon_s/E$ and $\gamma_l/E$.  
For droplets of all sizes,  microscopic behaviour near the contact line is fluid-like. 
At distances from the contact line much smaller than $\Upsilon_s/E$, the system takes the form of the Neumann triangle where the upward force of the droplet surface tension is balanced by the surface tensions of the substrate. 
Over distances much greater than $\Upsilon_s/E$ the effects of the substrate surface tension diminish, and the substrate response is elastic.

Our model, which considers the case $\Upsilon_{sl}=\Upsilon_{sv}$, should be naturally extendable to treat more general surface stresses. 
To maintain local force balance at the contact line, we expect that the surfaces will rearrange so that $\gamma_{lv}, \Upsilon_{sl} $ and $\Upsilon_{sv}$ still obey Neumann's triangle.
For large liquid surface tensions, the surface strain may be sufficiently large to reveal a strain-dependent contribution to $\Upsilon_s$. 
While we have been careful to avoid equating the surface stresses in a solid, $\Upsilon_{s}$, to the surface energy, $\gamma_{s}$, 
 equivalence of surface stress and surface energy has previously been seen experimentally in several solid materials \cite{peth57}.
Specifically, we expect $\Upsilon$ to be isotropic and equal to $\gamma$ for polymer gels \cite{jeri11} -- since their surfaces primarily consist of liquid solvent.

\begin{acknowledgments}
The authors gratefully acknowledge helpful conversations with Larry Wilen, John Wettlaufer and an anonymous referee. RWS is funded by the Yale University Bateman Interdepartmental Postdoctoral Fellowship.
\end{acknowledgments}

\appendix

\section{Calculating the macroscopic contact angle on a soft substrate}

\begin{figure}[h]
\centering
  \includegraphics[width=9cm]{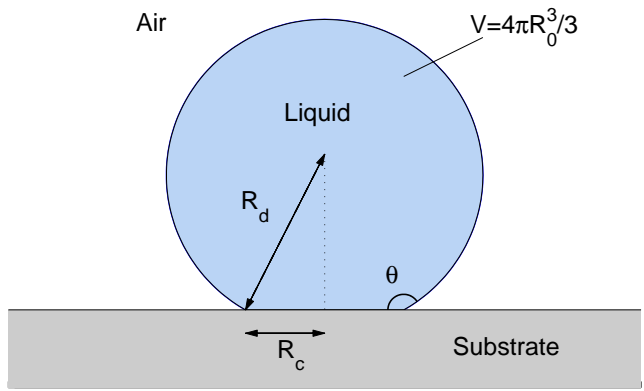}
  \caption{Schematic diagram for the contact angle calculation.}
  \label{fig:contact:schem}
\end{figure}

In order to calculate the contact angle on a soft substrate for $R\gtrsim \Upsilon_s/E$ we use the specific case of a droplet on an incompressible, thin substrate with $\gamma_{sl}=\gamma_{sv}=\gamma_s$. Thus $\nu=1/2$ and we assume $R\gg h$, as considered in Section \ref{sec:thin_substrate}. For a hemispherical droplet in this limit, the height of the peak is independent of $R$, and the pressure contribution to $F_{el}$ in Equation (\ref{eqn:Fel}) vanishes as $u_z(r,h)\rightarrow 0$ inside the droplet \cite{jeri11}, giving $F_{el}=(1.5 R\gamma_l^2/E) f(\Upsilon_s/Eh)$, where the function $f$ is the integral in Equation (\ref{eqn:R_gg_h}). When the macroscopic contact angle changes from $90^\circ$, then the vertical component of surface tension reduces to $\gamma_l \sin\theta$ and so $F_{el}\approx (1.5 R_c\gamma_l^2\sin^2\theta/E) f(\Upsilon_s/Eh)$, where $R_c$ is the radius of the contact area, shown in Figure \ref{fig:contact:schem}.

The surface energy contribution to the free energy can be calculated by assuming that the deflection of the substrate is small relative to the size of the droplet, so we have the scenario shown in Figure \ref{fig:contact:schem}. Then, from geometrical considerations, $F_{surf}=2\pi R_d^2\gamma_l(1-\cos\theta)+\pi R_c^2 (\gamma_{sl}-\gamma_{sv})$, where $R_c=R_d\sin\theta$. Assuming that the total volume of the droplet is fixed so that $V=4/3\pi R_0^3$ we find the equilibrium contact angle by numerically minimising $F=F_{surf}(V,\theta)+F_{el}(V,\theta)$. The results are then plotted in Figure \ref{fig:contact_angle_free_energy}(b) as a function of the free droplet radius $R_0$ divided by  $\Upsilon_s/E$. The figure also shows the contact angle for small droplets in the limit $R_0\ll \Upsilon_s/E$, where the macroscopic contact angle is determined by Neumann's triangle, so that $\theta=\cos^{-1}(\gamma_l/2\Upsilon_s)$.

\section{Fluid-like behaviour at the contact line}

Here, we demonstrate that in the absence of long-ranged intermolecular forces, the behaviour close to the contact line always reverts to the Neumann triangle at sufficiently small length scales. 
We start with the equation for the substrate profile under a droplet, Equation (\ref{eqn:uz}), and split the integral into two parts:
\begin{equation}
u_z(r,h)=\int_0^{E/\Upsilon_s} ds + \int^\infty_{E/\Upsilon_s}ds =u_z^{long}+u_z^{short}.
\end{equation}
Since contributions from each wavenumber, $s$, separately satisfy the governing equations, each of the two parts represents a valid elastic solution. 
The first integral represents the long wavelength ($\lambda>\Upsilon_s/E$) contribution to the surface profile, and is therefore smooth and cusp-free. 
The second integral represents the short wavelength contribution, and therefore contains all the details of the peak. 
For $r-R\ll \Upsilon_s/E,R$, we find that the second integral reduces to
\begin{equation}
u_z^{short}(r-R,h)=\frac{\gamma_l}{\Upsilon_s}\int_{E/\Upsilon_s}^{\infty}\frac{\cos{[s(r-R)]}}{\pi s^2}ds
\end{equation}
which has a symmetric peak of width ${\cal O}(\Upsilon_s/E)$ with slope $\pm\gamma_l/2\Upsilon_s$ either side.
 This is the Neumann triangle after linearisation for small surface gradients. Thus for regions of size $\ll O(\Upsilon_s/E,R)$ the contact line appears the same as a three-fluid contact line.


%


\begin{thebibliography}{33}%
\makeatletter
\providecommand \@ifxundefined [1]{%
 \@ifx{#1\undefined}
}%
\providecommand \@ifnum [1]{%
 \ifnum #1\expandafter \@firstoftwo
 \else \expandafter \@secondoftwo
 \fi
}%
\providecommand \@ifx [1]{%
 \ifx #1\expandafter \@firstoftwo
 \else \expandafter \@secondoftwo
 \fi
}%
\providecommand \natexlab [1]{#1}%
\providecommand \enquote  [1]{``#1''}%
\providecommand \bibnamefont  [1]{#1}%
\providecommand \bibfnamefont [1]{#1}%
\providecommand \citenamefont [1]{#1}%
\providecommand \href@noop [0]{\@secondoftwo}%
\providecommand \href [0]{\begingroup \@sanitize@url \@href}%
\providecommand \@href[1]{\@@startlink{#1}\@@href}%
\providecommand \@@href[1]{\endgroup#1\@@endlink}%
\providecommand \@sanitize@url [0]{\catcode `\\12\catcode `\$12\catcode
  `\&12\catcode `\#12\catcode `\^12\catcode `\_12\catcode `\%12\relax}%
\providecommand \@@startlink[1]{}%
\providecommand \@@endlink[0]{}%
\providecommand \url  [0]{\begingroup\@sanitize@url \@url }%
\providecommand \@url [1]{\endgroup\@href {#1}{\urlprefix }}%
\providecommand \urlprefix  [0]{URL }%
\providecommand \Eprint [0]{\href }%
\providecommand \doibase [0]{http://dx.doi.org/}%
\providecommand \selectlanguage [0]{\@gobble}%
\providecommand \bibinfo  [0]{\@secondoftwo}%
\providecommand \bibfield  [0]{\@secondoftwo}%
\providecommand \translation [1]{[#1]}%
\providecommand \BibitemOpen [0]{}%
\providecommand \bibitemStop [0]{}%
\providecommand \bibitemNoStop [0]{.\EOS\space}%
\providecommand \EOS [0]{\spacefactor3000\relax}%
\providecommand \BibitemShut  [1]{\csname bibitem#1\endcsname}%
\let\auto@bib@innerbib\@empty
\bibitem [{\citenamefont {de~Gennes}\ \emph {et~al.}(2010)\citenamefont
  {de~Gennes}, \citenamefont {Brochard-Wyart},\ and\ \citenamefont
  {Quere}}]{dege10}%
  \BibitemOpen
  \bibfield  {author} {\bibinfo {author} {\bibfnamefont {P.-G.}\ \bibnamefont
  {de~Gennes}}, \bibinfo {author} {\bibfnamefont {F.}~\bibnamefont
  {Brochard-Wyart}}, \ and\ \bibinfo {author} {\bibfnamefont {D.}~\bibnamefont
  {Quere}},\ }\href@noop {} {\emph {\bibinfo {title} {Capillarity and Wetting
  Phenomena: Drops, Bubbles, Pearls, Waves}}}\ (\bibinfo  {publisher}
  {Springer},\ \bibinfo {year} {2010})\BibitemShut {NoStop}%
\bibitem [{\citenamefont {Butt}\ \emph {et~al.}(2003)\citenamefont {Butt},
  \citenamefont {Graf},\ and\ \citenamefont {Kappl}}]{butt03}%
  \BibitemOpen
  \bibfield  {author} {\bibinfo {author} {\bibfnamefont {H.-J.}\ \bibnamefont
  {Butt}}, \bibinfo {author} {\bibfnamefont {K.}~\bibnamefont {Graf}}, \ and\
  \bibinfo {author} {\bibfnamefont {M.}~\bibnamefont {Kappl}},\ }\href@noop {}
  {\emph {\bibinfo {title} {Physics and Chemistry of Interfaces}}}\ (\bibinfo
  {publisher} {Wiley-VCH},\ \bibinfo {year} {2003})\ p.\ \bibinfo {pages}
  {153}\BibitemShut {NoStop}%
\bibitem [{\citenamefont {Quere}(2008)}]{quer08}%
  \BibitemOpen
  \bibfield  {author} {\bibinfo {author} {\bibfnamefont {D.}~\bibnamefont
  {Quere}},\ }\href@noop {} {\bibfield  {journal} {\bibinfo  {journal} {Ann.
  Rev. Mater. Res.}\ }\textbf {\bibinfo {volume} {38}},\ \bibinfo {pages} {71}
  (\bibinfo {year} {2008})}\BibitemShut {NoStop}%
\bibitem [{\citenamefont {Kumar}\ and\ \citenamefont {Prabhu}(2007)}]{kuma07}%
  \BibitemOpen
  \bibfield  {author} {\bibinfo {author} {\bibfnamefont {G.}~\bibnamefont
  {Kumar}}\ and\ \bibinfo {author} {\bibfnamefont {K.~N.}\ \bibnamefont
  {Prabhu}},\ }\href@noop {} {\bibfield  {journal} {\bibinfo  {journal} {Adv.
  Coll. Interf. Sci.}\ }\textbf {\bibinfo {volume} {133}},\ \bibinfo {pages}
  {61} (\bibinfo {year} {2007})}\BibitemShut {NoStop}%
\bibitem [{\citenamefont {Young}(1805)}]{youn05}%
  \BibitemOpen
  \bibfield  {author} {\bibinfo {author} {\bibfnamefont {T.}~\bibnamefont
  {Young}},\ }\href@noop {} {\bibfield  {journal} {\bibinfo  {journal} {Phil.
  Trans. R. Soc. London}\ }\textbf {\bibinfo {volume} {95}},\ \bibinfo {pages}
  {65} (\bibinfo {year} {1805})}\BibitemShut {NoStop}%
\bibitem [{\citenamefont {Neumann}(1894)}]{neum94}%
  \BibitemOpen
  \bibfield  {author} {\bibinfo {author} {\bibfnamefont {F.}~\bibnamefont
  {Neumann}},\ }\href@noop {} {\emph {\bibinfo {title} {Vorlesungen Ÿber die
  Theorie der CapillaritŠt}}}\ (\bibinfo  {publisher} {B.G. Teubner, Leipzig},\
  \bibinfo {year} {1894})\BibitemShut {NoStop}%
\bibitem [{\citenamefont {Michaels}\ and\ \citenamefont {Dean}(1962)}]{mich62}%
  \BibitemOpen
  \bibfield  {author} {\bibinfo {author} {\bibfnamefont {A.~S.}\ \bibnamefont
  {Michaels}}\ and\ \bibinfo {author} {\bibfnamefont {S.~W.}\ \bibnamefont
  {Dean}},\ }\href {\doibase 10.1021/j100816a005} {\bibfield  {journal}
  {\bibinfo  {journal} {The Journal of Physical Chemistry}\ }\textbf {\bibinfo
  {volume} {66}},\ \bibinfo {pages} {1790} (\bibinfo {year} {1962})},\ \Eprint
  {http://arxiv.org/abs/http://pubs.acs.org/doi/pdf/10.1021/j100816a005}
  {http://pubs.acs.org/doi/pdf/10.1021/j100816a005} \BibitemShut {NoStop}%
\bibitem [{\citenamefont {Shuttleworth}(1950)}]{shut50}%
  \BibitemOpen
  \bibfield  {author} {\bibinfo {author} {\bibfnamefont {R.}~\bibnamefont
  {Shuttleworth}},\ }\href {http://stacks.iop.org/0370-1298/63/i=5/a=302}
  {\bibfield  {journal} {\bibinfo  {journal} {Proceedings of the Physical
  Society. Section A}\ }\textbf {\bibinfo {volume} {63}},\ \bibinfo {pages}
  {444} (\bibinfo {year} {1950})}\BibitemShut {NoStop}%
\bibitem [{\citenamefont {Lester}(1961)}]{lest61}%
  \BibitemOpen
  \bibfield  {author} {\bibinfo {author} {\bibfnamefont {G.}~\bibnamefont
  {Lester}},\ }\href {\doibase 10.1016/0095-8522(61)90032-0} {\bibfield
  {journal} {\bibinfo  {journal} {J. Colloid Sci.}\ }\textbf {\bibinfo {volume}
  {16}},\ \bibinfo {pages} {315 } (\bibinfo {year} {1961})}\BibitemShut
  {NoStop}%
\bibitem [{\citenamefont {Jerison}\ \emph {et~al.}(2011)\citenamefont
  {Jerison}, \citenamefont {Xu}, \citenamefont {Wilen},\ and\ \citenamefont
  {Dufresne}}]{jeri11}%
  \BibitemOpen
  \bibfield  {author} {\bibinfo {author} {\bibfnamefont {E.~R.}\ \bibnamefont
  {Jerison}}, \bibinfo {author} {\bibfnamefont {Y.}~\bibnamefont {Xu}},
  \bibinfo {author} {\bibfnamefont {L.~A.}\ \bibnamefont {Wilen}}, \ and\
  \bibinfo {author} {\bibfnamefont {E.~R.}\ \bibnamefont {Dufresne}},\ }\href
  {\doibase 10.1103/PhysRevLett.106.186103} {\bibfield  {journal} {\bibinfo
  {journal} {Phys. Rev. Lett.}\ }\textbf {\bibinfo {volume} {106}},\ \bibinfo
  {pages} {186103} (\bibinfo {year} {2011})}\BibitemShut {NoStop}%
\bibitem [{\citenamefont {Leonforte}\ and\ \citenamefont
  {Muller}(2011)}]{leon11}%
  \BibitemOpen
  \bibfield  {author} {\bibinfo {author} {\bibfnamefont {F.}~\bibnamefont
  {Leonforte}}\ and\ \bibinfo {author} {\bibfnamefont {M.}~\bibnamefont
  {Muller}},\ }\href@noop {} {\bibfield  {journal} {\bibinfo  {journal} {J.
  Chem. Phys.}\ }\textbf {\bibinfo {volume} {135}},\ \bibinfo {pages} {214703}
  (\bibinfo {year} {2011})}\BibitemShut {NoStop}%
\bibitem [{\citenamefont {Eslami}\ and\ \citenamefont
  {Elliott}(2011)}]{esla11}%
  \BibitemOpen
  \bibfield  {author} {\bibinfo {author} {\bibfnamefont {F.}~\bibnamefont
  {Eslami}}\ and\ \bibinfo {author} {\bibfnamefont {J.~A.~W.}\ \bibnamefont
  {Elliott}},\ }\href@noop {} {\bibfield  {journal} {\bibinfo  {journal} {J.
  Phys. Chem. B}\ }\textbf {\bibinfo {volume} {115}},\ \bibinfo {pages} {10646}
  (\bibinfo {year} {2011})}\BibitemShut {NoStop}%
\bibitem [{\citenamefont {Mora}\ \emph {et~al.}(2010)\citenamefont {Mora},
  \citenamefont {Phou}, \citenamefont {Fromental}, \citenamefont {Pismen},\
  and\ \citenamefont {Pomeau}}]{mora10}%
  \BibitemOpen
  \bibfield  {author} {\bibinfo {author} {\bibfnamefont {S.}~\bibnamefont
  {Mora}}, \bibinfo {author} {\bibfnamefont {T.}~\bibnamefont {Phou}}, \bibinfo
  {author} {\bibfnamefont {J.-M.}\ \bibnamefont {Fromental}}, \bibinfo {author}
  {\bibfnamefont {L.~M.}\ \bibnamefont {Pismen}}, \ and\ \bibinfo {author}
  {\bibfnamefont {Y.}~\bibnamefont {Pomeau}},\ }\href {\doibase
  10.1103/PhysRevLett.105.214301} {\bibfield  {journal} {\bibinfo  {journal}
  {Phys. Rev. Lett.}\ }\textbf {\bibinfo {volume} {105}},\ \bibinfo {pages}
  {214301} (\bibinfo {year} {2010})}\BibitemShut {NoStop}%
\bibitem [{\citenamefont {Irgens}(2008)}]{irge08}%
  \BibitemOpen
  \bibfield  {author} {\bibinfo {author} {\bibfnamefont {F.}~\bibnamefont
  {Irgens}},\ }\href@noop {} {\emph {\bibinfo {title} {Continuum mechanics}}}\
  (\bibinfo  {publisher} {Springer-Verlag},\ \bibinfo {address} {Berlin
  Heidelberg},\ \bibinfo {year} {2008})\BibitemShut {NoStop}%
\bibitem [{\citenamefont {Sneddon}(1951)}]{sned51}%
  \BibitemOpen
  \bibfield  {author} {\bibinfo {author} {\bibfnamefont {I.~N.}\ \bibnamefont
  {Sneddon}},\ }\href@noop {} {\emph {\bibinfo {title} {Fourier Transforms}}}\
  (\bibinfo  {publisher} {McGraw-Hill},\ \bibinfo {year} {1951})\BibitemShut
  {NoStop}%
\bibitem [{\citenamefont {Xu}\ \emph {et~al.}(2010)\citenamefont {Xu},
  \citenamefont {Engl}, \citenamefont {Jerison}, \citenamefont {Wallenstein},
  \citenamefont {Hyland}, \citenamefont {Wilen},\ and\ \citenamefont
  {Dufresne}}]{xu10}%
  \BibitemOpen
  \bibfield  {author} {\bibinfo {author} {\bibfnamefont {Y.}~\bibnamefont
  {Xu}}, \bibinfo {author} {\bibfnamefont {W.~C.}\ \bibnamefont {Engl}},
  \bibinfo {author} {\bibfnamefont {E.~R.}\ \bibnamefont {Jerison}}, \bibinfo
  {author} {\bibfnamefont {K.~J.}\ \bibnamefont {Wallenstein}}, \bibinfo
  {author} {\bibfnamefont {C.}~\bibnamefont {Hyland}}, \bibinfo {author}
  {\bibfnamefont {L.~A.}\ \bibnamefont {Wilen}}, \ and\ \bibinfo {author}
  {\bibfnamefont {E.~R.}\ \bibnamefont {Dufresne}},\ }\href@noop {} {\bibfield
  {journal} {\bibinfo  {journal} {PNAS}\ }\textbf {\bibinfo {volume} {107}},\
  \bibinfo {pages} {14964} (\bibinfo {year} {2010})}\BibitemShut {NoStop}%
\bibitem [{\citenamefont {Pethica}\ and\ \citenamefont
  {Pethica}(1957)}]{peth57}%
  \BibitemOpen
  \bibfield  {author} {\bibinfo {author} {\bibfnamefont {B.~A.}\ \bibnamefont
  {Pethica}}\ and\ \bibinfo {author} {\bibfnamefont {T.~J.~P.}\ \bibnamefont
  {Pethica}},\ }in\ \href@noop {} {\emph {\bibinfo {booktitle} {Proc. 2nd Int.
  Cong. Surface Activity}}},\ Vol.~\bibinfo {volume} {3}\ (\bibinfo {year}
  {1957})\BibitemShut {NoStop}%
\bibitem [{\citenamefont {Cammarata}\ and\ \citenamefont
  {Sieradzki}(1994)}]{camm94}%
  \BibitemOpen
  \bibfield  {author} {\bibinfo {author} {\bibfnamefont {R.}~\bibnamefont
  {Cammarata}}\ and\ \bibinfo {author} {\bibfnamefont {K.}~\bibnamefont
  {Sieradzki}},\ }\href@noop {} {\bibfield  {journal} {\bibinfo  {journal}
  {Ann. Rev. Mater. Sci.}\ }\textbf {\bibinfo {volume} {24}},\ \bibinfo {pages}
  {215} (\bibinfo {year} {1994})}\BibitemShut {NoStop}%
\bibitem [{\citenamefont {Spaepen}(2000)}]{spae00}%
  \BibitemOpen
  \bibfield  {author} {\bibinfo {author} {\bibfnamefont {F.}~\bibnamefont
  {Spaepen}},\ }\href@noop {} {\bibfield  {journal} {\bibinfo  {journal} {Acta
  mater.}\ }\textbf {\bibinfo {volume} {48}},\ \bibinfo {pages} {31} (\bibinfo
  {year} {2000})}\BibitemShut {NoStop}%
\bibitem [{\citenamefont {Marchand}\ \emph {et~al.}(2012)\citenamefont
  {Marchand}, \citenamefont {Das}, \citenamefont {Snoeijer},\ and\
  \citenamefont {Andreotti}}]{marc12}%
  \BibitemOpen
  \bibfield  {author} {\bibinfo {author} {\bibfnamefont {A.}~\bibnamefont
  {Marchand}}, \bibinfo {author} {\bibfnamefont {S.}~\bibnamefont {Das}},
  \bibinfo {author} {\bibfnamefont {J.~H.}\ \bibnamefont {Snoeijer}}, \ and\
  \bibinfo {author} {\bibfnamefont {B.}~\bibnamefont {Andreotti}},\ }\href@noop
  {} {\bibfield  {journal} {\bibinfo  {journal} {Phys. Rev. Lett.}\ }\textbf
  {\bibinfo {volume} {108}},\ \bibinfo {pages} {094301} (\bibinfo {year}
  {2012})}\BibitemShut {NoStop}%
\bibitem [{\citenamefont {Long}\ \emph {et~al.}(1996)\citenamefont {Long},
  \citenamefont {Ajdari},\ and\ \citenamefont {Leibler}}]{long96}%
  \BibitemOpen
  \bibfield  {author} {\bibinfo {author} {\bibfnamefont {D.}~\bibnamefont
  {Long}}, \bibinfo {author} {\bibfnamefont {A.}~\bibnamefont {Ajdari}}, \ and\
  \bibinfo {author} {\bibfnamefont {L.}~\bibnamefont {Leibler}},\ }\href@noop
  {} {\bibfield  {journal} {\bibinfo  {journal} {Langmuir}\ }\textbf {\bibinfo
  {volume} {12}},\ \bibinfo {pages} {5221} (\bibinfo {year}
  {1996})}\BibitemShut {NoStop}%
\bibitem [{\citenamefont {Roman}\ and\ \citenamefont {Bico}(2010)}]{roma10}%
  \BibitemOpen
  \bibfield  {author} {\bibinfo {author} {\bibfnamefont {B.}~\bibnamefont
  {Roman}}\ and\ \bibinfo {author} {\bibfnamefont {J.}~\bibnamefont {Bico}},\
  }\href@noop {} {\bibfield  {journal} {\bibinfo  {journal} {J. Phys.: Condens.
  Matter}\ }\textbf {\bibinfo {volume} {22}},\ \bibinfo {pages} {493101}
  (\bibinfo {year} {2010})}\BibitemShut {NoStop}%
\bibitem [{\citenamefont {Shanahan}\ and\ \citenamefont
  {de~Gennes}(1986)}]{shan86}%
  \BibitemOpen
  \bibfield  {author} {\bibinfo {author} {\bibfnamefont {M.}~\bibnamefont
  {Shanahan}}\ and\ \bibinfo {author} {\bibfnamefont {P.-G.}\ \bibnamefont
  {de~Gennes}},\ }\href@noop {} {\bibfield  {journal} {\bibinfo  {journal} {C.
  R. Acad. Sc. Paris}\ }\textbf {\bibinfo {volume} {302}},\ \bibinfo {pages}
  {517} (\bibinfo {year} {1986})}\BibitemShut {NoStop}%
\bibitem [{\citenamefont {Pericet-Cámara}\ \emph {et~al.}(2008)\citenamefont
  {Pericet-Cámara}, \citenamefont {Best}, \citenamefont {Butt},\ and\
  \citenamefont {Bonaccurso}}]{peri08}%
  \BibitemOpen
  \bibfield  {author} {\bibinfo {author} {\bibfnamefont {R.}~\bibnamefont
  {Pericet-Cámara}}, \bibinfo {author} {\bibfnamefont {A.}~\bibnamefont
  {Best}}, \bibinfo {author} {\bibfnamefont {H.-J.}\ \bibnamefont {Butt}}, \
  and\ \bibinfo {author} {\bibfnamefont {E.}~\bibnamefont {Bonaccurso}},\
  }\href {\doibase 10.1021/la801862m} {\bibfield  {journal} {\bibinfo
  {journal} {Langmuir}\ }\textbf {\bibinfo {volume} {24}},\ \bibinfo {pages}
  {10565} (\bibinfo {year} {2008})},\ \bibinfo {note} {pMID: 18720996},\
  \Eprint {http://arxiv.org/abs/http://pubs.acs.org/doi/pdf/10.1021/la801862m}
  {http://pubs.acs.org/doi/pdf/10.1021/la801862m} \BibitemShut {NoStop}%
\bibitem [{\citenamefont {Carre}\ \emph {et~al.}(1996)\citenamefont {Carre},
  \citenamefont {Gastel},\ and\ \citenamefont {Shanahan}}]{carr96}%
  \BibitemOpen
  \bibfield  {author} {\bibinfo {author} {\bibfnamefont {A.}~\bibnamefont
  {Carre}}, \bibinfo {author} {\bibfnamefont {J.-C.}\ \bibnamefont {Gastel}}, \
  and\ \bibinfo {author} {\bibfnamefont {M.~E.~R.}\ \bibnamefont {Shanahan}},\
  }\href@noop {} {\bibfield  {journal} {\bibinfo  {journal} {Nature}\ }\textbf
  {\bibinfo {volume} {379}},\ \bibinfo {pages} {432} (\bibinfo {year}
  {1996})}\BibitemShut {NoStop}%
\bibitem [{\citenamefont {Pericet-Camara}\ \emph {et~al.}(2009)\citenamefont
  {Pericet-Camara}, \citenamefont {Auernhammer}, \citenamefont {Koynov},
  \citenamefont {Lorenzoni}, \citenamefont {Raiteri},\ and\ \citenamefont
  {Bonaccurso}}]{peri09}%
  \BibitemOpen
  \bibfield  {author} {\bibinfo {author} {\bibfnamefont {R.}~\bibnamefont
  {Pericet-Camara}}, \bibinfo {author} {\bibfnamefont {G.~K.}\ \bibnamefont
  {Auernhammer}}, \bibinfo {author} {\bibfnamefont {K.}~\bibnamefont {Koynov}},
  \bibinfo {author} {\bibfnamefont {S.}~\bibnamefont {Lorenzoni}}, \bibinfo
  {author} {\bibfnamefont {R.}~\bibnamefont {Raiteri}}, \ and\ \bibinfo
  {author} {\bibfnamefont {E.}~\bibnamefont {Bonaccurso}},\ }\href {\doibase
  10.1039/B907212H} {\bibfield  {journal} {\bibinfo  {journal} {Soft Matter}\
  }\textbf {\bibinfo {volume} {5}},\  (\bibinfo {year} {2009})}\BibitemShut
  {NoStop}%
\bibitem [{\citenamefont {White}(2003)}]{whit03}%
  \BibitemOpen
  \bibfield  {author} {\bibinfo {author} {\bibfnamefont {L.~R.}\ \bibnamefont
  {White}},\ }\href@noop {} {\bibfield  {journal} {\bibinfo  {journal} {J.
  Colloid Interface Sci.}\ }\textbf {\bibinfo {volume} {258}},\ \bibinfo
  {pages} {82} (\bibinfo {year} {2003})}\BibitemShut {NoStop}%
\bibitem [{\citenamefont {Soon}\ and\ \citenamefont {Mu}(1986)}]{soon86}%
  \BibitemOpen
  \bibfield  {author} {\bibinfo {author} {\bibfnamefont {H.~Y.}\ \bibnamefont
  {Soon}}\ and\ \bibinfo {author} {\bibfnamefont {S.~J.}\ \bibnamefont {Mu}},\
  }\href@noop {} {\bibfield  {journal} {\bibinfo  {journal} {J. Colloid
  Interface Sci.}\ }\textbf {\bibinfo {volume} {110}},\ \bibinfo {pages} {252}
  (\bibinfo {year} {1986})}\BibitemShut {NoStop}%
\bibitem [{\citenamefont {Sokuler}\ \emph {et~al.}(2010)\citenamefont
  {Sokuler}, \citenamefont {Auernhammer}, \citenamefont {Roth}, \citenamefont
  {Liu}, \citenamefont {Bonacurrso},\ and\ \citenamefont {Butt}}]{soku10}%
  \BibitemOpen
  \bibfield  {author} {\bibinfo {author} {\bibfnamefont {M.}~\bibnamefont
  {Sokuler}}, \bibinfo {author} {\bibfnamefont {G.~K.}\ \bibnamefont
  {Auernhammer}}, \bibinfo {author} {\bibfnamefont {M.}~\bibnamefont {Roth}},
  \bibinfo {author} {\bibfnamefont {C.}~\bibnamefont {Liu}}, \bibinfo {author}
  {\bibfnamefont {E.}~\bibnamefont {Bonacurrso}}, \ and\ \bibinfo {author}
  {\bibfnamefont {H.-J.}\ \bibnamefont {Butt}},\ }\href {\doibase
  10.1021/la903996j} {\bibfield  {journal} {\bibinfo  {journal} {Langmuir}\
  }\textbf {\bibinfo {volume} {26}},\ \bibinfo {pages} {1544} (\bibinfo {year}
  {2010})},\ \bibinfo {note} {pMID: 19928793},\ \Eprint
  {http://arxiv.org/abs/http://pubs.acs.org/doi/pdf/10.1021/la903996j}
  {http://pubs.acs.org/doi/pdf/10.1021/la903996j} \BibitemShut {NoStop}%
\bibitem [{\citenamefont {Shanahan}(1987)}]{shan87}%
  \BibitemOpen
  \bibfield  {author} {\bibinfo {author} {\bibfnamefont {M.}~\bibnamefont
  {Shanahan}},\ }\href@noop {} {\bibfield  {journal} {\bibinfo  {journal} {J.
  Phys. D: Appl. Phys.}\ }\textbf {\bibinfo {volume} {20}},\ \bibinfo {pages}
  {945} (\bibinfo {year} {1987})}\BibitemShut {NoStop}%
\bibitem [{\citenamefont {Das}\ \emph {et~al.}(2011)\citenamefont {Das},
  \citenamefont {Marchand}, \citenamefont {Andreotti},\ and\ \citenamefont
  {Snoeijer}}]{das11}%
  \BibitemOpen
  \bibfield  {author} {\bibinfo {author} {\bibfnamefont {S.}~\bibnamefont
  {Das}}, \bibinfo {author} {\bibfnamefont {A.}~\bibnamefont {Marchand}},
  \bibinfo {author} {\bibfnamefont {B.}~\bibnamefont {Andreotti}}, \ and\
  \bibinfo {author} {\bibfnamefont {J.~H.}\ \bibnamefont {Snoeijer}},\
  }\href@noop {} {\bibfield  {journal} {\bibinfo  {journal} {Phys. Fluids}\
  }\textbf {\bibinfo {volume} {23}},\ \bibinfo {pages} {072006} (\bibinfo
  {year} {2011})}\BibitemShut {NoStop}%
\bibitem [{\citenamefont {Maxwell}(1878)}]{maxw78}%
  \BibitemOpen
  \bibfield  {author} {\bibinfo {author} {\bibfnamefont {J.~C.}\ \bibnamefont
  {Maxwell}},\ }\enquote {\bibinfo {title} {Encylopaedia britannica},}\ \
  (\bibinfo  {publisher} {Samuel L. Hall, New York},\ \bibinfo {year} {1878})\
  Chap.\ \bibinfo {chapter} {Capillary Action}, p.~\bibinfo {pages}
  {56}\BibitemShut {NoStop}%
\bibitem [{\citenamefont {Olives}(2010)}]{oliv10}%
  \BibitemOpen
  \bibfield  {author} {\bibinfo {author} {\bibfnamefont {J.}~\bibnamefont
  {Olives}},\ }\href@noop {} {\bibfield  {journal} {\bibinfo  {journal} {J.
  Phys.: Condens. Matter}\ }\textbf {\bibinfo {volume} {22}},\ \bibinfo {pages}
  {085005} (\bibinfo {year} {2010})}\BibitemShut {NoStop}%
\end{thebibliography}
\end{document}